\documentclass[amsmath,amssymb,aip,jcp,twocolumn,10pt]{revtex4-1}

\usepackage{amsmath,amssymb}
\usepackage{dcolumn}
\usepackage{graphicx}
\usepackage{algpseudocode,algorithm}
\usepackage[usenames]{color}

\begin{document}

\title{Molecular Cluster Perturbation Theory. I. Formalism}


\author{Jason N. Byrd}
\email{byrdja@chem.ufl.edu}
\affiliation{Quantum Theory Project, University of Florida, Gainesville, FL 32611}
\author{Nakul Jindal}
\affiliation{Department of Computer Science and Engineering, University of Florida, Gainesville, FL 32611}
\author{Robert W. {Molt, Jr.}}
\affiliation{Quantum Theory Project, University of Florida, Gainesville, FL 32611}
\author{Rodney J. Bartlett}
\affiliation{Quantum Theory Project, University of Florida, Gainesville, FL 32611}
\author{Beverly A. Sanders}
\affiliation{Department of Computer Science and Engineering, University of Florida, Gainesville, FL 32611}
\author{Victor F. Lotrich}
\affiliation{Quantum Theory Project, University of Florida, Gainesville, FL 32611}


\begin{abstract}
We present second-order molecular cluster perturbation theory (MCPT(2)), a linear scaling methodology to calculate arbitrarily large systems with explicit calculation of individual wavefunctions in a coupled-cluster framework.  This new MCPT(2) framework uses coupled-cluster perturbation theory and an expansion in terms of molecular dimer interactions to obtain molecular wavefunctions that are infinite-order in both the electronic fluctuation operator and all possible dimer (and products of dimers) interactions.  The MCPT(2) framework has been implemented in the new SIA/{Aces4} parallel architecture, making use of the advanced dynamic memory control and fine grained parallelism to perform very large explicit molecular cluster calculations.  To illustrate the power of this method, we have computed energy shifts, lattice site dipole moments, and harmonic vibrational frequencies via explicit calculation of the bulk system for the polar and non-polar polymorphs of solid hydrogen fluoride.  The explicit lattice size (without using any periodic boundary conditions) was expanded up to 1,000 HF molecules, with 32,000 basis functions and 10,000 electrons.  Our obtained HF lattice site dipole moments and harmonic vibrational frequencies agree well with the existing literature.
\end{abstract}
\maketitle

\section{Introduction}

There are at least two fundamental challenges in modern quantum chemistry: the
efficient calculation of both dynamic and static correlation simultaneously, and
the calculation of successively larger systems.  Diverse classes of methods
exist \cite{flocke2004,tomasi2005, gordon2007, fedorov2012, klamt2011,
vreven2006, podeszwa2008, mayhall2011,knizia2012} to describe large systems
within a quantum mechanical framework by subdivision into smaller fragments.
For an overarching summary of modern fragmentation methods, we refer to the
review article of Gordon {\it et al.} \cite{gordon2012} In general, when
presented with a large number of degrees of freedom, one has the choice to
represent them implicitly or explicitly.  Examples of implicit representation
include employing periodic boundary conditions or approximating the surrounding
system as a bath with a continuum representation.  Explicit representations of
high numbers of degrees of freedom typically require sub-partitioning of the
system  in some manner to reduce the computational cost.  Our goal in this work
is to create an explicit method of handling large systems with a rigorously
defined perturbation expansion.
We consider it highly desirable to have an explicit model. Inherent to 
periodic boundary conditions or implicit solvation models is a 
requirement of homogeneity. Inherently heterogeneous phenomena, such as 
crystalline defects, solvents with varied electrical moments over the 
molecule (like carboxylic acids), or surface interactions require some 
explicit representation of these degrees of freedom.


If we consider an arbitrarily large system consisting of $N$ non-interacting
(infinitely separated) molecular clusters, collectively referred to as monomers,
we can write the Schr\"{o}dinger equation for each monomer $X$ as 
\begin{equation} \label{monS1}
\hat{H}_X |\phi_X\rangle = E_X |\phi_X\rangle
\end{equation}
where $|\phi_X\rangle$ is the properly anti-symmetric monomer wavefunction.  The
total { system's} wavefunction is then a simple direct product of the monomer
wavefunctions
\begin{equation}\label{prodspace}
|\Psi_p\rangle = |\phi_1\rangle\otimes|\phi_2\rangle\otimes\dots|\phi_N\rangle.
\end{equation}
Assuming that these monomers retain their identity to a large degree as they
begin to interact (by adiabatically bringing them to some finite separation),
then we can preserve the product nature of the system wavefunction by
introducing the complete anti-symmetrizer $\hat{A}$ so that
\begin{equation}\label{prodwf}
|\Psi\rangle = 
\hat{A}|\Psi_p\rangle =
\hat{A}|\phi_1\rangle\otimes|\phi_2\rangle\otimes|\phi_3\rangle\otimes\dots|\phi_N\rangle.
\end{equation}
The anti-symmetrizer $\hat{A}$ can be written in terms of the permutation
operator $\hat{P}$ as 
\begin{equation}
\hat{A}=1+\hat{P}
\end{equation}
where $\hat{P}$ acting on an electronic wavefunction interchanges electrons
between monomers in all possible ways resulting in a completely anti-symmetric
wavefunction.  This product wavefunction is a common Ansatz in molecular and
condensed matter physics distinguished by the various treatments of the
anti-symmetrizer and Hamiltonian approximations.  By further assuming that each
monomer interacts non-covalently with neighboring monomers, the problem becomes
an excellent candidate for an intermolecular force expansion
\cite{buckingham1967}.  Using the product
wavefunction Eq. \ref{prodspace} as our monomer fermi vacuum, with the
specification that the monomer wavefunctions $|\phi_X\rangle$ are the
Hartree-Fock solutions to Eq.  \ref{monS1}, we can write the system electronic
Hamiltonian without further approximation as the sum of individual monomer and
dimer Hamiltonians,
\begin{equation}  \label{clusterH}
\hat{H} = \sum^N_{X} \hat{H}_X + \sum^N_{X,Y\ne X} \hat{H}_{XY},
\end{equation}  
where $N$ is the total number of monomers.  This exact separation of the
Hamiltonian into only monomer and dimer spaces arises from the precisely block
diagonal nature of the product wavefunction in Eq. \ref{prodspace}, due to
approximating the global anti-symmetrizer as $\hat{A}\simeq 1$. 
Neglecting the global anti-symmetrizer in Eq.
\ref{prodwf} like this leads to the well known polarization expansion, whereas the term
including $\hat{P}$ leads to the exchange expansion, the details of
both expansions are well-documented \cite{rybak1991,jeziorski1994}. 

Traditional methods based on Eq. \ref{prodwf} 
(such as polarization \cite{chalasinski1977,jeziorski1994} or symmetry-adapted
perturbation theory \cite{rybak1991,jeziorski1994}) compute the wavefunction and
energy components perturbatively from these two expansions in terms
of polarization, induction, dispersion, exchange, and ``mixed" terms (see Rybak
{\it et al.} \cite{rybak1991} for a comprehensive discussion).  The standard approaches to the
intermolecular force perturbation expansion suffer from several significant
drawbacks; the expansion in terms of the intermonomer electrostatic operator is
nonconvergent and requires explicit inclusion of many-body intermonomer effects
\cite{lotrich1997,cvitas2007}. Also absent in most standard intermolecular expansions is an
explicit monomer wavefunction.  This lack of a monomer wavefunction greatly hampers
the evaluation of solvation shifted or bulk limit properties.  It is our
goal in this work to address some of the limitations found in traditional
methods by developing a monomer centric wavefunction theory that self
consistently includes the perturbative effects of the { surrounding} system.

The success of coupled-cluster (CC) theory \cite{bartlett2007} as a rapidly
converging description of dynamic correlation makes it a natural candidate for
molecular cluster interactions.  The behavior of CC theory is the driving force
behind many embedding methods
\cite{flocke2004,cammi2009,mayhall2011,bygrave2012,list2014}.  Such molecular
clusters are dominated by weaker intermolecular forces and are therefore the
most amenable to an intermolecular expansion.  We present a new infinite-order
pairwise-based embedding framework, which we refer to as molecular cluster
perturbation theory (MCPT), as a means to efficiently compute the
Schr\"{o}dinger equation for large systems of interacting molecular clusters.
The MCPT method combines the coupled-cluster perturbation theory
\cite{bartlett2010} (CCPT) with the intermolecular force approach
\cite{buckingham1967,jeziorski1994} of products of monomer wavefunctions.  By
expanding the perturbation series in terms of both the intramolecule (monomer)
and intermolecular (dimer) electron fluctuation operators, a formally consistent
perturbation theory can be obtained that has many significant advantages.
\begin{enumerate}

\item 
Expanding in terms of pairwise interactions reduces the extremely high
computational cost of { $O(N^6o^2v^4)$} inherent to the standard coupled-cluster
theory with all singles and doubles \cite{purvis1982} (CCSD) to { $O(N^2 o^2v^4)$
where $o$ ($v$) is the number of occupied (virtual) orbitals of an individual
monomer} while $N$ is the total number of monomers.  Further physical arguments
based on the distance scaling of intermolecular forces allow the introduction of
a cutoff radius past which all explicit interactions are neglected.  This cutoff
reduces the MCPT computational scaling to be linear with respect to the number
of monomers.

\item
Choosing the particle excitation rank partitioning
\cite{grabouski2007,bartlett2010} of the Hamiltonian
($\hat{H}=\hat{H}_0+\hat{V}$), the single and
double excitation cluster operators (monomer and dimer) completely decouple at
first-order in $\hat{V}$ while remaining infinite order in $\hat{H}_0$.

\item 
By our choice of $H_0$, all intermolecular interactions are
iterated over so that the final monomer wavefunction contains all possible
pairwise interactions and products of pairwise interactions to infinite-order.
This alleviates the non-convergent nature found in other intermolecular force
theories.

\item
We neglect the global anti-symmetrization of the total system's wavefunction by
approximating it as a product state of individual monomer
wavefunctions.  


\item The explicit monomer wavefunction nature of the MCPT formalism allows the
further computation of monomer-only properties while retaining the information
of the surrounding system.  Knowledge of the monomer wavefunctions allows the 
computation of any quantum mechanical observable (e.g. optical spectra) and the shifts to the observable 
due to the surrounding system (such as solvation shifts or crystal field effects.

\end{enumerate}
Due to the greatly improved computational scaling, the MCPT framework allows for
the calculation of large systems cheaply and explicitly without any continuum
representation or periodic boundary conditions.  In this work, we report the
study of 1,000 hydrogen fluoride molecules in two different crystal polymorphs.
This is one of the largest explicit quantum calculations on record, done on
merely $\sim256$ processors within a 12 hour queue. 

The HF lattice is one of the simplest systems to describe physically due to the
small number of electrons and non-covalent interaction between HF molecules. In
the field of calculating correlated wavefunctions for crystalline systems, new
methods need simple systems against which to compare to understand the merits of
new many-body methodologies.  We consider the HF lattice to be a good candidate
as a prototype benchmark system.
The HF lattice is additionally of theoretical interest as it is not currently
known what crystal polymorph is most stable between polar and non-polar forms
\cite{ atoji1954, johnson1975, otto1986, panas1993, berski1998, buth2004,
buth2006, sode2010, bygrave2012}.  Work toward this goal is an interesting
challenge for new methods, given that previous estimates suggest that the
difference between polymorphs at a few kcal/mol. The computational vibrational
spectra also poses a challenge for new correlated wavefunction crystal methods
as the crystal induced shifts of the IR and Raman spectra are tremendous
\cite{kittelberger1967,anderson1980,pinnick1989,sode2009,sode2012}.

This manuscript is organized as follows:  In Sec. \ref{mcptsec2} a brief
overview of coupled-cluster perturbation theory and the particle rank
partitioning of $\hat{H}_0$ is presented.  This is followed by Sec.
\ref{mcptsec3} where the derivation and use of the effective molecular cluster
Hamiltonians in the CCPT framework is summarized.  Working MCPT(2) spin-adapted
equations, with other supporting equations, are given in Sec. \ref{mcptsec4}.
After a brief summary of the electronic structure calculations in Sec.
\ref{struc}, we present in Sec. \ref{calccrystal} our large scale calculations
on the $10\times10\times10$ square HF crystal.  Sec. \ref{conclusions} contains
our concluding remarks.

\section{\label{mcptsec1}Molecular Cluster Perturbation Theory}

\subsection{\label{mcptsec2}Second-order coupled cluster perturbation theory and choice of $\hat{H}_0$} 

The CCPT expansion is outlined in detail in this section so that we can build
upon the formal definitions later to construct the MCPT equations. 
For a generic system the Schr{\"o}dinger equation can be written in terms of the 
coupled-cluster (CC) exponential expansion as, 
\begin{equation} \label{ccE1}
\hat{H} e^{\hat{T}} |\phi_0\rangle = E e^{\hat{T}} |\phi_0\rangle
\end{equation} 
where the cluster operator, $\hat{T}$, acting on the reference $|\phi_0\rangle$
creates $n$-fold excited determinants.  The Hamiltonian, $\hat{H}$, is given in
normal-ordered second-quantized form as
\begin{align}\label{Nhamiltonian}
\hat{H} = &
\langle 0|\hat{H}|0\rangle +
\sum_{pq}f^p_q \lbrace \hat{p}^\dag \hat{q}\rbrace +
\frac{1}{4}\sum_{pqrs}\bar{v}^{pq}_{rs} \lbrace \hat{p}^\dag \hat{q}^\dag
\hat{s}\hat{r}\rbrace \\
= &
\langle 0|\hat{H}|0\rangle +
\hat{F} + \hat{W}
\end{align}
where $f$ is the usual one particle Fock matrix, $\bar v$ are anti-symmetric
two-electron integrals and $\lbrace\rbrace$ denotes normal ordering of the
included operators.  
The Hamiltonian can be formally partitioned as a sum of operators $\hat{H}_0$
and $\hat{V}$, where 
\begin{equation}\label{partH1}
\hat{H}_0 |\phi_0\rangle = E_0 |\phi_0\rangle
\end{equation} 
and 
\begin{equation}
\hat{V}\equiv\hat{H}-\hat{H}_0
\end{equation} 
is treated as a perturbation.  Writing Eq. \ref{ccE1} and the partitioned
Hamiltonian from Eq. \ref{partH1} we have
\begin{equation}  
\left(\hat{H}_0 + \hat{V}\right) e^{\hat{T}} |\phi_0\rangle = E
e^{\hat{T}}|\phi_0\rangle.
\end{equation}  
From here it is possible to obtain the cluster amplitudes and final energy through
projection against the appropriate excitation space.  Performing the standard similarity
transformation,
\begin{equation}  
e^{-\hat{T}} \left(\hat{H}_0 + \hat{V}\right) e^{\hat{T}} |\phi_0\rangle = 
E |\phi_0\rangle,
\end{equation}  
the energy is obtained by projection with the reference space
\begin{equation}  
\langle\phi_0|e^{-\hat{T}} \left(\hat{H}_0 + \hat{V}\right) e^{\hat{T}}
|\phi_0\rangle = E \langle\phi_0|\phi_0\rangle
\end{equation}  
while the cluster amplitudes are defined by projecting with the excited
manifold $|\phi^g\rangle\langle\phi^g|$ giving
\begin{equation}  
\langle\phi^g|e^{-\hat{T}} \left(\hat{H}_0 + \hat{V}\right) e^{\hat{T}} |\phi_0\rangle = 0  
\end{equation}  
where $\langle\phi^g|$ indicates a $g$-fold excited determinant.
The cluster operator $\hat{T}$ can be expanded in the perturbation $\hat{V}$, leading to 
the CCPT equations for the first-order amplitudes and second-order energy:
\begin{equation}  \label{ccpt2wf}
\langle\phi^g|\hat{V} + \left[\hat{H}_0, \hat{T}\right] |\phi_0\rangle = 0  
\end{equation}  
\begin{equation}  \label{ccpt2e}
\langle\phi_0|\left[\hat{V}, \hat{T}\right] |\phi_0\rangle = E^{(2)}.
\end{equation}  

\begin{figure}
\center\includegraphics[width=0.5\textwidth]{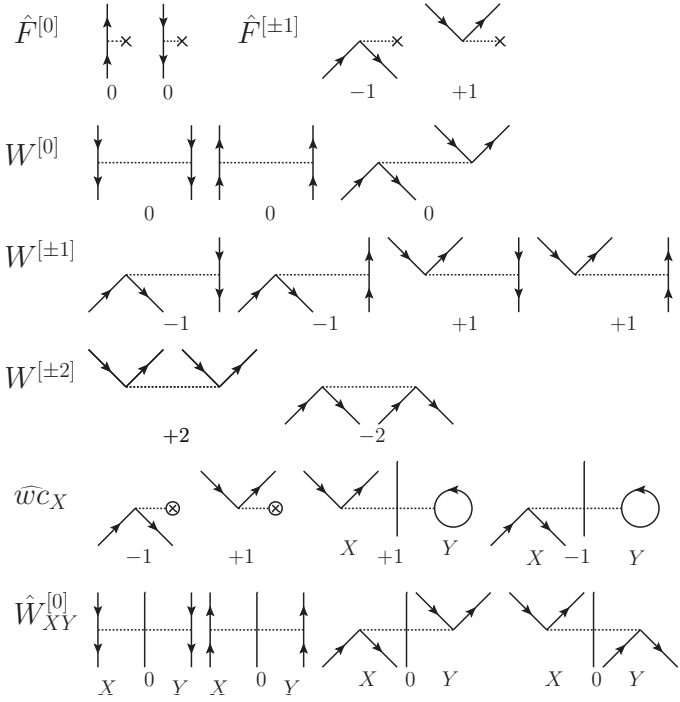}
\caption{\label{diagram}Diagrammatic form of the one- $\hat{F}^{[n]}$ and
two-particle $\hat{W}^{[n]}$ (Eq. \ref{Nhamiltonian}) operators with particle
excitation rank number $n$ listed.  Also shown are the effective one-electron
$\widehat{wc}_X$ (Eq. \ref{wc}) and dimer $W^{[0]}_{XY}$ Brandow-type diagrams
(the remaining dimer $W^{[n]}_{XY}$ are expressed in a similar fashion, see Eq.
\ref{NHxy} for the algebraic form).
} \end{figure}

It is customary to choose $\hat{H}_0$ to be the one-particle Fock
operator $\hat{F}$, commonly referred to as the M{\o}ller-Plesset partitioning.  This
leads to the standard MPn equations \cite{shavitt2009} which have been
quintessential to modern computational chemistry. However, other choices in partitioning
of the Hamiltonian are possible.  In this work we will use the particle rank partitioning,
\begin{equation} \label{partition1}
\hat{H}_0 = \hat{F}^{[0]} + \hat{W}^{[0]},
\end{equation} 
used in our previous work \cite{bartlett2010,byrd2014-b} where $[0]$ denotes
only particle excitation rank conserving contributions.  This choice has the
advantages over the MPn series that at second-order in the energy (Eq.
\ref{ccpt2e}), the singles and doubles amplitude equations from Eq.
\ref{ccpt2wf} are linear in the cluster operator $\hat{T}$ and completely
decoupled from each other (as contrasted with the standard CCSD theory, which
introduces significant coupling between the singles and doubles cluster
amplitudes).  From this definition of $\hat{H}_0$, the perturbation 
\begin{equation}\label{partition2}
\hat{V}=\hat{H}-\hat{H}_0 = \hat{F}^{[\pm 1]} + \hat{W}^{[\pm 1,\pm 2]}
\end{equation}
will contain all terms that do not preserve particle excitation rank
(see Fig. \ref{diagram} for the diagrammatic forms of the $\hat{F}^{[n]}$ and
$\hat{W}^{[n]}$ operators).

If we use canonical Hartree-Fock orbitals in our reference wavefunction
$|\phi_0\rangle$, with the choice of $\hat{H}_0$ defined above, the first-order
amplitudes and second-order energy are equivalent to LCCD (linear
coupled-cluster theory).  However, as will be illustrated below, for
non-Hartree-Fock orbitals this LCCD amplitude and energy equivalence to CCPT(2)
is no longer valid.  When applied brute-force to chemical systems, CCPT(2) is
faster than the traditional CCSD due to the removed CPU and I/O cost of
computing and storing the quadratic product terms.  It is a Hermitian theory,
which is theoretically more straight forward in the determination of properties.
Its Hermitian nature also allows for the calculation of properties (including
analytic gradients) twice as quickly as compared to standard coupled-cluster
theory, as computation of the Lambda equations is avoided.  

\subsection{\label{mcptsec3}Effective molecular cluster Hamiltonians}

Before proceeding to the { derivation} of second-order MCPT, it is convenient
to define the following monomer specific index domains
\begin{equation}
p,p_1,r,r_1,a,i,a_1,i_1,\dots\in {\cal H}(X)
\end{equation}
and
\begin{equation}
q,q_1,s,s_1,b,j,b_1,j_1,\dots\in {\cal H}(Y)
\end{equation}
where $a,b,\dots$ refer to virtual orbitals and $i,j,\dots$ refer to occupied
orbitals while $p,q,\dots$ range over both occupied and virtual orbitals.  These
index domains will be used to completely define the range of all tensors used
hereafter.  The monomer Hamiltonian, $\hat{H}_X$, is then given in
second-quantized form by
\begin{align}\label{NHx}
\hat{H}_X =  &
\langle 0|\hat{H}_X|0\rangle +
\sum_{pp_1}f^p_{p_1} \lbrace \hat{p}^\dag \hat{p}_1\rbrace +
\frac{1}{4}\sum_{pp_1rr_1}\bar{v}^{pp_1}_{rr_1} \lbrace 
\hat{p}^\dag \hat{p}^\dag_1 \hat{r}_1\hat{r}\rbrace \\
= &
\langle 0|\hat{H}_X|0\rangle +
\hat{F}_X + \hat{W}_X
\end{align} 
while the dimer Hamiltonian $\hat{H}_{XY}$ is given by
\begin{align}\label{NHxy}
\hat{H}_{XY} =  &
\langle 0|\hat{H}_{XY}|0\rangle 
+ 
\sum_{pp_1}h^p_{p_1} \lbrace \hat{p}^\dag \hat{p}_1\rbrace  \\ &
+ \sum_{qq_1}h^q_{q_1} \lbrace \hat{q}^\dag \hat{q}_1\rbrace
+ \sum_{pqrs}\bar{v}^{pq}_{rs} \lbrace \hat{p}^\dag \hat{q}^\dag \hat{s}\hat{r}\rbrace \\
= &
\langle 0|\hat{H}_{XY}|0\rangle +
\widehat{wc}^{[\pm 1]}_{X} + \widehat{wc}^{[\pm 1]}_Y 
+ \hat{W}_{XY}.
\end{align} 
Here, the monomer and dimer one- ($\hat{F}_X$) and two- ($\hat{W}_X$ and
$\hat{W}_{XY}$) particle operators are defined appropriately in either the
monomer or dimer subspace as dictated by the indices.  
The one-particle operator $\widehat{wc}_X$
(see Fig.  \ref{diagram}) is defined \cite{rybak1991} as
\begin{equation}\label{wc}
\widehat{wc}_X=w^a_i \hat{a}^\dag\hat{i} 
\end{equation}
where the matrix elements
\begin{equation}
w^a_i = h^a_i + 2 v^{aj}_{ij},
\end{equation}
\begin{widetext}
are composed of the one-electron integrals
\begin{equation}
h^a_i = 
-\sum_{\mu\nu\in{\cal H}(X)} C_{\mu a}C_{\nu i}
\int d{\bf r_1}d{\bf r_2}
\psi^*_{\mu}({\bf r_1}) 
\sum_{\substack{Y\ne X\\ \alpha\in Y}} \frac{Z_\alpha}{R_\alpha-r_1}
\psi_{\nu}({\bf r_1})
\end{equation}
which includes the effect of all system nuclei on the wavefunction of $X$, where
$C_{\mu p}$ are the monomer Hartree-Fock coefficients and  $\psi_{\nu}({\bf r_1})$ 
are gaussian basis functions belonging to monomer $X$.  The $v^{aj}_{ij}$ two-electron integrals 
are evaluated as a special case from 
\begin{equation}\label{dp2e}
(pr|qs) = 
\sum_{\substack{
    \mu\nu\in{\cal H}(X) \\
    \lambda\sigma\in{\cal H}(Y)}}
    C_{\mu p}C_{\nu r}
    C_{\lambda q}C_{\sigma s}
\int d{\bf r_1}d{\bf r_2}
\psi^*_{\mu}({\bf r_1}) \psi_{\nu}({\bf r_1}) 
r^{-1}_{12}
\psi^*_{\lambda}({\bf r_2}) \psi_{\sigma}({\bf r_2})
\end{equation}
\end{widetext}
with $\lambda=\sigma$.  It should be noted that the Hartree-Fock orbitals on
each monomer are not orthogonalized with other monomers.
Using the approximation of Eq. \ref{prodspace} to define our Hartree-Fock vacuum, we can develop our MCPT embedding method.

The monomer and dimer Hamiltonians can still be partitioned in terms of
reference and perturbation Hamiltonians
\begin{equation}  
\hat{H}_X =  \hat{H}_{X0} + \hat{V}_{X}
\end{equation}  
and 
\begin{equation}
\hat{H}_{XY} =  \hat{H}_{XY0} + \hat{V}_{XY},
\end{equation}
where we again use the particle excitation rank partitioning given by (see
Fig. \ref{diagram} for the Brandow-type diagrams for $\hat{H}_{XY0}$)
\begin{align}
\hat{H}_{X0} & = \hat{F}^{[0]}_X + \hat{W}^{[0]}_X, \\
\hat{V}_{X}  & = \hat{F}^{[\pm 1]}_X + \hat{W}^{[\pm 1]}_X + \hat{W}^{[\pm 2]}_X, \\
\hat{H}_{XY0}& =  \hat{W}^{[0]}_{XY}, \text{~and} \\
\hat{V}_{XY}& =  \widehat{wc}^{[\pm 1]}_X + \widehat{wc}^{[\pm 1]}_Y 
+ \hat{W}^{[\pm 1]}_{XY} + \hat{W}^{[\pm 2]}_{XY}.
\end{align}
Throughout this work, the lower case ``$t$" is
reserved for singles amplitudes, with upper case ``$T$" and script ``$\cal T$" 
reserved for monomer and dimer doubles amplitudes, respectively.  
Using this notation the system coupled-cluster amplitudes can be expressed as 
\begin{equation}  \label{clusterT}
\hat{T} = \sum_X \hat{T}_X + \sum_{X,Y\ne X} \hat{T}_{XY}
\end{equation}  
where the cluster operator for monomer $X$, 
\begin{equation}\label{monomerT}
\hat{T}_X = {\hat t}^{(1)}_X + {\hat T}^{(2)}_X,
\end{equation}
contains the singles and doubles operators defined as
\begin{equation} \label{1mt1}
{\hat t}^{(1)}_X = \sum_{ai} t^a_i \hat{a}^{\dagger} \hat{i}
\end{equation}  
and
\begin{equation}  \label{1mt2}
{\hat T}^{(2)}_X  = \frac{1}{4}\sum_{aa_1ii_1} T^{a a_1}_{i i_1} \lbrace\hat{a}^{\dagger} \hat{a}_1^{\dagger}
\hat{i}_1 \hat{i}\rbrace,
\end{equation}  
respectively, while the cluster operator for the $XY$ dimer only has a doubles
contribution (there can be no ${\hat t}^{(1)}_{XY}$ in a monomer centered
basis) given by
\begin{equation}  \label{1dt2}
\hat{T}_{XY} =\hat{T}^{(2)}_{XY} = \sum_{abij}{\cal T}^{a b}_{i j} \hat{a}^{\dagger} \hat{b}^{\dagger} \hat{j}
\hat{i}.
\end{equation}  

Using Eqs. \ref{clusterT}, \ref{monomerT}, and \ref{1dt2} it is possible to
transform the system Hamiltonian (Eq.  \ref{clusterH}) into an effective monomer
and dimer Hamiltonian as
\begin{equation}\label{meffH1}
{\tilde H}_X =
\left[
e^{-\hat{T}+\hat{T}_X} \hat{H} e^{\hat{T}-\hat{T}_X} 
\right]_{X} 
\end{equation}
and
\begin{equation}\label{deffH1}
{\tilde H}_{XY} = 
\left[
e^{-\hat{T}+\hat{T}^{(2)}_{XY}} \hat{H} e^{\hat{T}-\hat{T}^{(2)}_{XY}} 
\right]_{XY}
\end{equation}
where the brackets $\left[\dots\right]_X$ and $\left[\dots\right]_{XY}$ denote
that only terms operating in the monomer ${\cal H}(X)$ and dimer ${\cal
H}(X)\otimes {\cal H}(Y)$ Hilbert spaces remain.  All terms within the brackets
not a member of the appropriate final space to be are internally contracted away.  Expanding the
Baker-Campbell-Hausdorff (BCH) commutator in Eq.
\ref{meffH1} in the same manner as Eqs. \ref{ccpt2wf} and \ref{ccpt2e} (keeping
to the appropriate order in $\hat{V}$) gives
\begin{align}
\tilde{H}_X & =
\hat{H}_X + \widehat{wc}_X + 
\sum_{Y\ne X}\left[\hat{H}_{XY0},
\hat{T}_{Y} +
\hat{T}_{XY}\right] \\
& = \hat{H}_{X0} + \tilde{V}_X
\end{align}  
with
\begin{equation}\label{vtwiddle}
\tilde{V}_X = 
\hat{V}_X + \widehat{wc}_X +
\sum_{Y\ne X}
\left[\hat{H}_{XY0},\hat{t}^{(1)}_{Y}+\hat{T}^{(2)}_{XY}\right].
\end{equation}
Similarly, the dimer effective Hamiltonian in Eq. \ref{deffH1} can be expanded to
give
\begin{equation}
\tilde{H}_{XY} = \hat{H}_{XY0} +  \hat{V}_{XY} + 
\left[ \hat{H}_{XY0}, \hat{T}^{(2)}_{X} + \hat{T}^{(2)}_{Y} \right].
\end{equation}
These effective Hamiltonians have the property that the monomer
and dimer Hamiltonians are at this point only members of that specific Hilbert space:
$\tilde{H}_X\in{\cal H}(X)$ and $\tilde{H}_{XY}\in\lbrace{\cal H}(X)\otimes{\cal
H}(Y)={\cal H}(XY)\rbrace$.  Contributions from $Y$ in the $\tilde{H}_X$
effective Hamiltonian are completely internally contracted away at this point.
The monomer and dimer polarization components of the
Schr\"{o}dinger equation for the system can now be written, respectively, as
\begin{equation}\label{effS1}
e^{-\hat{T}_X}
\left(\tilde{H}_{X} - E^{(2)}_{\rm monomer}\right)
e^{\hat{T}_X}|\phi_X\rangle = 0
\end{equation}
and
\begin{equation}\label{effS2}
e^{-\hat{T}_{XY}}
\left(\tilde{H}_{XY} - E^{(2)}_{\rm dimer}\right)
e^{\hat{T}_{XY}}|\phi_X\rangle\otimes|\phi_Y\rangle = 0.
\end{equation}
\begin{widetext}
The corresponding amplitude equations can be obtained straightforwardly
from Eqs. \ref{effS1} and \ref{effS2} giving
\begin{equation}\label{mcpt-Mwf}
\langle\phi_X^g|
\left[\hat{H}_{X0},\hat{T}_X\right] 
+ \hat{V}_X + \widehat{wc}_X 
+ 
\sum_{Y\ne X}\left[\hat{H}_{XY0},\hat{t}^{(1)}_{Y}+\hat{T}^{(2)}_{XY}\right]
|\phi_X\rangle = 0.
\end{equation}
and
\begin{equation}\label{mcpt-Dwf}
\langle\phi_Y^g|\otimes
\langle\phi_X^g|
\left[\hat{H}_{XY0},\hat{T}_{XY}\right]
+\hat{V}_{XY} 
+  
\left[ \hat{H}_{XY0}, \hat{T}^{(2)}_{X} + \hat{T}^{(2)}_{Y} \right]
|\phi_X\rangle\otimes |\phi_Y\rangle = 0.
\end{equation}
\end{widetext}

By transforming to these effective Hamiltonians, we have made clear the 
electrostatic, induction and dispersion-like monomer and
dimer effects from monomer $Y$ into the monomer $X$ Hamiltonian.  However,
instead of the physically { separable} bare integral terms \cite{rybak1991}
these contributions are inseparably included to infinite-order through the
coupled-cluster amplitudes by the terms
\begin{equation}
\widehat{wc}_X +
\sum_{Y\ne X}\left[\hat{H}_{XY0},\hat{t}^{(1)}_{Y}+\hat{T}^{(2)}_{XY}\right].
\end{equation}
There are also the corresponding monomer corrections into the dimer potential
via
\begin{equation}
\left[ \hat{H}_{XY0}, \hat{T}^{(2)}_{X} + \hat{T}^{(2)}_{Y} \right].
\end{equation}
The inclusion of the induction/dispersion interaction of all other monomers,
$Y$, in the amplitude equation of monomer $X$ through Eq. \ref{vtwiddle} has
several important implications.  Firstly, at zeroth-order each monomer
wavefunction includes the inductive field of the surrounding system.  Secondly,
while interactions are limited to pairwise terms only, each monomer feeds back into
the inductive field of the system iteratively.  This means that each monomer
includes not just all pairwise interactions but all products of pairwise
interactions to infinite-order.  To graphically illustrate this point, Fig.
\ref{hflatticeplot}c contains an illustration of the pairwise communication
topology.  Here monomers $A$, $B$ and $C$ and monomers in the $D'$ and $E'$
grouping have all combinations of pairwise interactions at zeroth-order.  On the
second step of the iterative solution monomer $A$ will now contain information
from the entire $D'$ and $E'$ grouping as well as the $BC$ interaction.
Continuing this process iteratively it is evident that, for example, monomer $A$
will contain the information of all possible pairwise interactions.  An
approximation of immediate concern is the use of a cutoff radius ($R_{\rm cut}$)
when deciding what explicit dimer interactions to include in equations
\ref{mcpt-Mwf} and \ref{mcpt-Dwf}.  The introduction of such a cut off greatly
reduces the computational cost of the amplitude equations, while introducing
what would be a small error due to the infinite-order nature of the pairwise
contribution.

\subsection{\label{mcptsec4}Working spin-adapted MCPT(2) equations and program flow}

Standard second quantization techniques can be used to obtain the cluster and
pairwise additive amplitudes as well as the associated monomer and 2-body
polarization energy equations.  For the sake of brevity, we forgo a detailed
derivation and simply present the final spin-adapted equations using the
spin-restricted Hartree-Fock reference as implemented in our program.  With the
cluster definitions from Eqs. \ref{monomerT}-\ref{1dt2} and the energy Eqs.
\ref{effS1} and \ref{effS2} we can write the explicit energy equations.  From
here on summation over repeated lower and upper indices is implied.  The monomer
$X$ singles (M1) and doubles (M2) energy contributions are given by 
\begin{equation}\label{em1}
E^{(2)}_{\rm M1}(X) = 
2 w^i_a t^a_i   
\end{equation} 
and
\begin{equation} \label{em2}
E^{(2)}_{\rm M2}(X) = 
(2v^{i i_1}_{a a_1}-v^{i_1 i}_{a a_1}) T^{a a_1}_{i i_1}
\end{equation} 
for the doubles energy while the $XY$ dimer doubles (D2) energy is defined as
\begin{equation} \label{ed2}
E^{(2)}_{\rm D2}(XY) = 4 v^{ij}_{ab} {\cal T}^{ab}_{ij}.
\end{equation} 
The monomer doubles energy {\it shift} is relative to the isolated monomer $X$ LCCD energy
as 
\begin{equation}\label{dem2}
\delta E^{(2)}_{\rm M2}(X) = 
E^{(2)}_{\rm M2}(X) - E_{\rm LCCD}(X).
\end{equation}
Using Eqs. \ref{monomerT}-\ref{1dt2} with Eqs. \ref{mcpt-Mwf} and \ref{mcpt-Dwf}
the monomer singles amplitude equations are 
\begin{equation} \label{tm1}
\epsilon^i_a t^a_i 
+ (2v^{ai_1}_{ia_1} - v^{ai}_{a_1 i_1})t^{a_1}_{i_1} 
+ w^a_i + 
\sum_{\substack{Y\ne X_A\\ c,k\in {\cal H}(Y)}}
\Bigl(v^{ak}_{ic} t^c_k \Bigr) = 0,
\end{equation} 
\begin{widetext}
while the monomer and dimer doubles amplitudes are defined as
\begin{multline} \label{tm2}
\epsilon^{ii_1}_{aa_1} T^{a a_1}_{i i_1} 
+ v^{a a_1}_{i i_1} 
+ v^{a a_1}_{a_2 a_3} T^{a_2 a_3}_{i i_1} 
+ v_{i i_1}^{i_2 i_3} T^{a a_1}_{i_2 i_3}  
+ (v^{a i_2}_{i a_2} - v^{a i_2}_{a_2 i}) T^{a_2 a_1}_{i_2 i_1} 
+ (v^{a_1 i_2}_{i_1 a_2} - v^{a_1 i_2}_{a_2 i_1}) T^{a_2 a}_{i_2 i} \\
+ v^{i_2 a}_{a_2 i} (T^{a_2 a_1}_{i_2 i_1} - T^{a_2 a_1}_{i_1 i_2}) 
+ v^{i_2 a_1}_{a_2 i_1} (T^{a_2 a}_{i_2 i} - T^{a_2 a}_{i i_2})   
- v^{a_1 i_2}_{a_2 i}T^{a a_2}_{i_2 i_1}  
- v^{a i_2}_{a_2 i_1}T^{a_1 a_2}_{i_2 i} \\
+ \sum_{\substack{Y\ne X_A\\ c,k\in {\cal H}(Y)}}
\left(v^{a_1 k}_{i_1 c} {\cal T}^{ac}_{ik} +  v^{a k}_{i c}
{\cal T}^{a_1 c}_{i_1 k}\right) = 0 
\end{multline} 
and
\begin{multline} 
\label{td2}
\epsilon^{ij}_{ab} {\cal T}^{ab}_{ij}
+ v^{ab}_{ij} 
+ v^{ab}_{a_1 b_1} {\cal T}^{a_1 b_1}_{ij} 
+ v^{i_1 j_1}_{ij} {\cal T}^{ab}_{i_1 j_1}
- v^{b i_1}_{b_1 i} {\cal T}^{a b_1}_{i_1 j} 
- v^{a j_1}_{a_1 j} {\cal T}^{a_1 b}_{i j_1}  
+ (2v^{i_1 a}_{a_1 i} - v^{a i_1}_{a_1 i}){\cal T}^{a_1 b}_{i_1 j} \\
+ (2v^{j_1 b}_{b_1 j} - v^{b j_1}_{b_1 j}){\cal T}^{a b_1}_{i j_1} 
+ \left( v^{i_1 b}_{a_1 j} (2{T}^{a a_1}_{i i_1} - {T}^{a a_1}_{i_1 i})
+ v^{j_1 a}_{b_1 i} (2T^{b b_1}_{j j_1} - T^{b b_1}_{j_1 j})\right)
= 0 
\end{multline} 
respectively.  
\end{widetext}
For clarity, the summation over the $c,k$ indices is explicitly written in
Eqs. \ref{tm1} and \ref{tm2}.  The energy tensor used above is defined as
\begin{equation}
\epsilon^{ij\cdots}_{ab\cdots} = 
\epsilon_i + \epsilon_j + \cdots - \epsilon_a - \epsilon_b - \cdots
\end{equation}
where $\epsilon_p$ is the $p$'th Hartree-Fock orbital energy, the energy denominator is then 
\begin{equation}
D^{ij\cdots}_{ab\cdots} = 1/\epsilon^{ij\cdots}_{ab\cdots}.
\end{equation}

The energy equations \ref{em1}-\ref{ed2}, and amplitude equations
\ref{tm1}-\ref{td2} define the second-order MCPT embedding method.  With all the
pertinent equations derived we are able now to summarize the overall algorithm.
Note that in the following $X$ and $Y$ remain dummy indices over monomers, $N$
is the number of monomers and $R_{\rm cut}$ is the distance cutoff value.
\begin{enumerate}
\item
For each monomer in the system a Hartree-Fock calculation is performed in the
monomer centered basis set neglecting all environment contributions.  The
resulting monomer specific molecular orbitals (MO) are stored.

\item 
The two-electron integrals are computed and transformed into the monomer, and
direct product MO basis (see Eq. \ref{dp2e}).  
If a cutoff radius is to be used, only dimers that satisfy the distance criteria
are included.

\item
The one-electron $\widehat{wc}$ operator is formed.

\item
Using $w^a_i$ as an initial guess to $t^a_i$ the singles amplitudes
are iterated over until convergence as: 
\begin{algorithmic}
\ForAll{CC iterations}
\For{$1\le X\le N$}
\ForAll{$Y$ satisfying $R_{\rm cut}$}
\State Construct $\sum v^{ak}_{ic} t^c_k$.
\EndFor
\State Construct $t^a_i$ (Eq. \ref{tm1}).
\State Compute $E^{(2)}_{\rm M1}$ (Eq. \ref{em1}).
\If{$E^{(2)}_{\rm M1}$ is converged}
\State Exit
\EndIf
\EndFor
\EndFor
\end{algorithmic}

\item
Compute LCCD energies for each monomer in the monomer centered basis set again
neglecting all environment contributions.

\item \label{doublesloop}
Using $v^{aa_1}_{ii_1}$ and $v^{ab}_{ij}$ as an initial guess, the 
$T^{aa_1}_{ii_1}$ and ${\cal T}^{ab}_{ij}$ doubles amplitudes are iterated
over in a macro and micro iteration fashion until converged as illustrated:
\begin{algorithmic}
\ForAll{CC macro iterations}
\For{$1\le X\le N$}
\ForAll{CC micro iterations}
\ForAll{$Y$ satisfying $R_{\rm cut}$} 
\State Construct 
$v^{a_1 k}_{i_1 c} {\cal T}^{ac}_{ik} +  v^{a k}_{i c} {\cal T}^{a_1 c}_{i_1 k}$
\EndFor
\State Construct $T^{aa_1}_{ii_1}$ (Eq. \ref{tm2}).
\State Compute $E^{(2)}_{\rm M2}$ (Eq. \ref{em2}).
\If{$E^{(2)}_{\rm M2}$ is converged}
\State Exit
\EndIf
\EndFor
\EndFor
\For{$1\le X\le N$, $1\le Y\le N$}
\If{$X\ne Y$ and $R_{\rm cut}$ is satisfied}
\ForAll{CC micro iterations}
\State Construct ${\cal T}^{ab}_{ij}$ (Eq. \ref{td2}).
\State Compute $E^{(2)}_{\rm M2}$ (Eq. \ref{ed2}).
\If{$E^{(2)}_{\rm D2}$ is converged}
\State Exit
\EndIf
\EndFor
\EndIf
\EndFor
\EndFor
\end{algorithmic}

\end{enumerate}

{ The practical computational benefit of using the MCPT(2) method can be seen by
examining the most intensive part of the calculation (Eqs. \ref{tm2} and
\ref{td2}) described in bullet
(\ref{doublesloop}) above.  Using $M$ and $\tilde{M}$ to denote the total number
of monomer- and dimer-doubles (first and second sub loops respectively) micro
iterations the effective computational scaling of the entire calculation is
$O\left((M+\tilde{M})\times N\tilde{N}\times o^2 v^4\right)$, where $N$ denotes
the number of monomers in the system, $\tilde{N}$ is the largest number of
monomers contained within the cutoff radius $R_{\rm cut}$ and the number of
occupied and virtual orbitals corresponds to the largest included monomer.  This
is contrasted by the corresponding LCCD scaling of $O(M\times N^6\times o^2
v^4)$ which will always be more costly so long as $\tilde{M}\tilde{N}<N^5$.  It
is our experience in practice for $\tilde{M}\sim15$ so that only for two
monomers is LCCD less computationally intensive. }

As an additional note, in the computation of first-order properties it is convenient
to work with the one-particle density matrix (1DM).
This monomer 1DM is defined in the AO basis as
\begin{equation}\label{1pdm}
\rho_{\mu\nu} = 
\sum_{ij} C_{\mu i} C_{\nu j}
-\sum_{pq}C_{\mu p} \rho_{pq} C_{\nu q}
\end{equation}
where $\rho_{pq}$ is the response
density computed from the monomer cluster amplitudes $\hat{T}_{X}$ and the
indices $p,q$ range over both occupied and virtual orbitals.
The contribution from exchange between monomers $X$ and $Y$ can be obtained at the
Hartree-Fock level by examining
\begin{equation}
\langle\phi_X|\hat{O}\hat{P}|\phi_X\rangle
\end{equation}
where $\cal O$ is the one-particle property operator, and we define then the Hartree-Fock exchange density as
\begin{equation}\label{1pdmexch}
\bar{\rho}_{\mu\bar{\nu}} = 
\sum_{\mu'\bar{\nu}'}
\sum_{ij} C_{\mu i}C_{\mu'j} 
S^{\bar{\nu}'}_{\mu'}
\sum_{\bar{k}\bar{\ell}} C_{\bar{\nu}'\bar{k}}C_{\bar{\nu}\bar{\ell}}.
\end{equation}
for all $\mu,\mu',i,j\in {\cal H}(X)$ and
$\bar{\nu},\bar{\nu}',\bar{k},\bar{\ell}\in {\cal H}(Y)$,
where $S^{\bar{\nu}'}_{\mu'}$ is the overlap matrix between
monomers \cite{rybak1991}.
A first-order property can be evaluated through the expectation value of the
wavefunction as 
\begin{equation}\label{1pprop}
\langle \hat{\cal O}\rangle = 
Tr({\bf {\cal O}\rho}) + Tr({\bf {\cal O\bar{\rho}}}),
\end{equation}
where the first term is the monomer only value and the second is the
non-local dimer contribution.

\section{\label{struc}Electronic Structure Calculations}

\begin{figure*}
\center\includegraphics[width=0.9\textwidth]{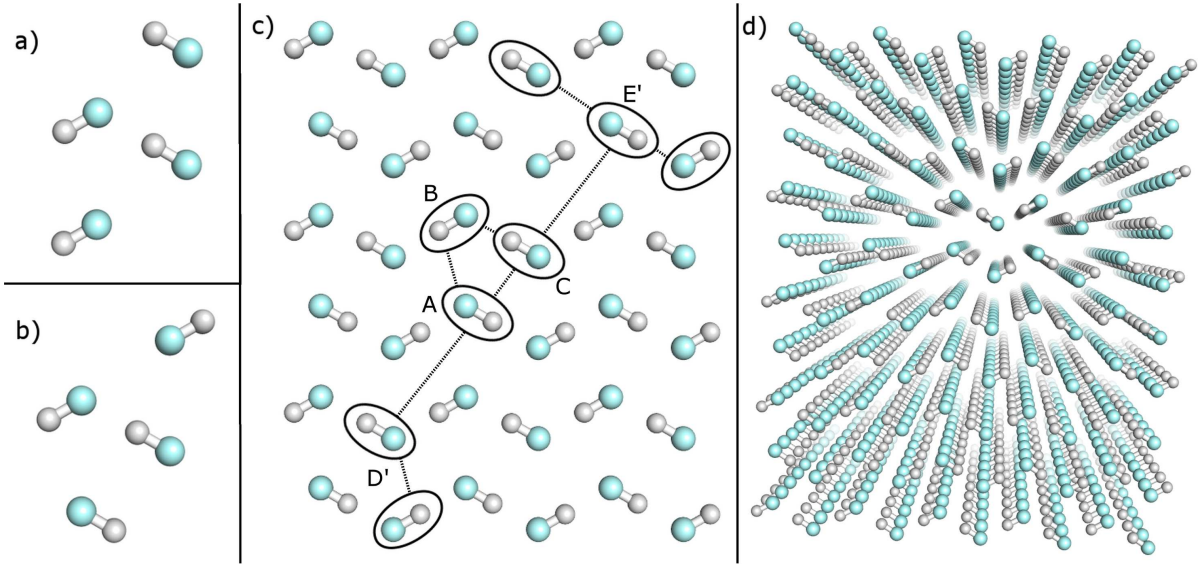}
\caption{\label{hflatticeplot}
HF molecules in a crystal;
a$)$ Polar unit cell,
b$)$ Non-Polar unit cell.
c$)$ Mono-layer of crystal HF in the non-polar ($8\times8\times8$) polymorph.  Also
shown diagrammatically is an example of the pairwise monomer-monomer
communication topology.
d$)$ Schematic of the largest HF crystal cube (show is the non-polar
polymorph) considered in this paper.  With 1,000 explicit HF molecules in
the aug-cc-pCVDZ basis set there is 32,000 basis set functions in total.
}
\end{figure*}

Electronic structure calculations were performed on the University of Florida
HiPerGator high performance cluster and the Cray (XE6) Garnet based at the ERDC DoD
Supercomputer Resource Center.  The all {\it ab initio} results were obtained
using the new {Aces4} massively parallel {\it ab initio} quantum chemistry
package based on the new implementation of the Super Instruction
Architecture \cite{lotrich2010} (SIA)\footnote{The new SIA framework development
page is hosted at https://github.com/UFParLab.} program.

The current implementation of MCPT does not include the ability to freeze the
core electrons of the monomers.  To guarantee a balanced treatment of the
electronic correlation, we use the core-valence version of the Dunning
correlation consistent basis sets \cite{woon1995} (cc-pCVnZ) on all heavy atoms
with the corresponding standard basis \cite{dunning1989} (cc-pVnZ) on hydrogen.
To aid in the basis set convergence of computed energies and dipole
moments \cite{halkier1999}, diffuse functions \cite{kendall1992} (aug-) were added
to each atom's basis set.  It is established in this work and elsewhere that
dipole moments \cite{halkier1999} and HF molecular harmonic frequencies are
described reasonably with just the aug-cc-pCVDZ basis set.


Immensely instrumental in the practical implementation and use of the MCPT
methodology is the recently reimplemented SIA.  The extremely large arrays
required are automatically partitioned and distributed, and are allocated only
as needed.  They can be very conveniently manipulated from
SIAL \cite{lotrich2010}, the domain specific programming language used to script
calculations.  This means that by expanding in the direct product space (Eq.
\ref{prodspace}) we have introduced a block sparsity into the system which is
fully exploited by the dynamic memory management system.

\section{Numerical Results and Discussion}

\subsection{\label{calccrystal}Hydrogen fluoride crystal}

\begin{figure}
\center\includegraphics[width=0.45\textwidth]{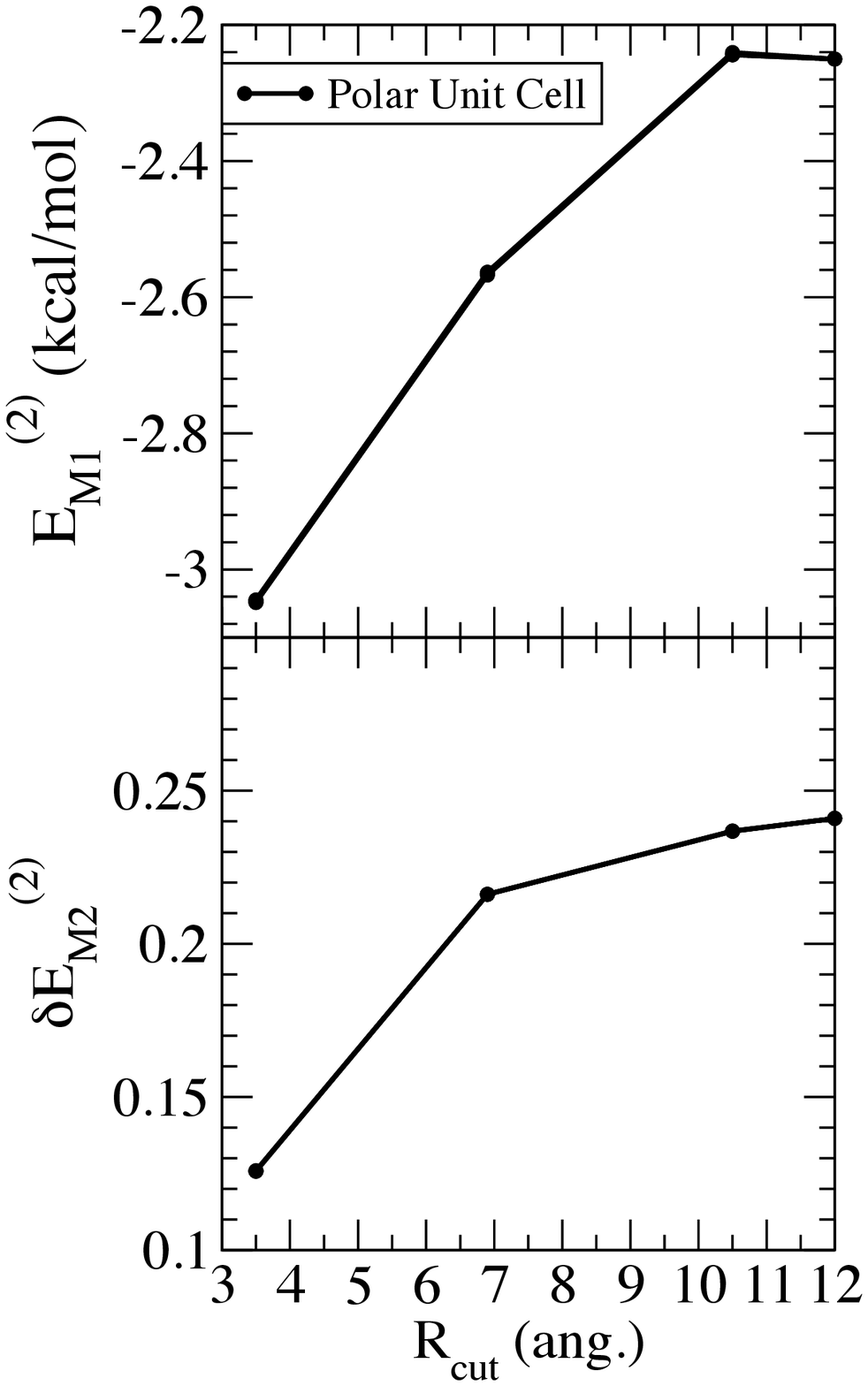}
\caption{\label{rcutscan}Convergence of the polar unit cell monomer energy shifts
as a function of the distance cutoff $R_{\rm cut}$.  The difference between the
singles energy shift at $10.5$ and $12.0$ \AA~is $4$ cm${}^{-1}$, and for the
doubles energy shift the difference is $1$ cm${}^{-1}$.  The cutoff values were
chosen to correspond to the separation between HF layers.}
\end{figure}

As previously discussed, the tremendous computational scaling reduction obtained
by working within an iterative pairwise interaction framework means that scaling
quantum systems to the bulk limit is accessible.  In this section, we illustrate
just how flexible this framework is by examining the HF molecular crystal from a
starting seed of eight molecules up through the bulk limit using a thousand
molecules.  The double $\zeta$ quality basis set (aug-cc-pCVDZ) was used here
for all calculations, which provides qualitatively accurate energetics (except
for very small energy differences on the order of a few kcal/mol) and well
converged dipole moments.  The total HF crystal structure was generated by
building up HF "1-D" polymers into a regular cube.  Spacings between HF
molecules within and between the polymer chains was chosen as to replicate the
experimental lattice constants \cite{atoji1954}.  The H-F bond length was fixed
at 0.92 \AA~ so as to agree with this the same experiment.  The two basic unit
cells, denoted polar and non-polar here, are illustrated in Fig.
\ref{hflatticeplot}a-b.  Also shown in Fig.  \ref{hflatticeplot}c-d is the HF
crystal cube mono-layer and the final 3-D structure.  Bulk limit scaling
calculations were performed for both the polar (Fig. \ref{hflatticeplot}a) and
non-polar (Fig. \ref{hflatticeplot}b) polymorphs.  The size of the HF cube for
both polymorphs was increased from $2\times2\times2$ to $10\times10\times10$,
systematically.  The largest structure considered (the $10\times10\times10$
lattice) has $\sim32,000$ basis set functions and $10,000$ electrons.  In all
crystal calculations a dimer radius cutoff of $12$ \AA~ was used.  
{ With these lattice parameters the size of the $10\times10\times10$
    crystal is $20\times20\times30$~{\AA}.  Choosing a $12$ \AA~ cutoff results
in a maximum of $\tilde{N}\sim350$ (with an average of $\tilde{N}\sim210$)
monomers included within the threshold.}  Numerical experimentation showed that
a much smaller ({next} nearest neighbors) cutoff is sufficient to converge the
MCPT doubles amplitudes to within {0.1 kcal/mol}.  However the electrostatic
contributions are not converged to the same degree until at least the $10.5$
\AA~ cutoff distance.  {This is illustrated for the polar lattice in Fig.  \ref{rcutscan}.}

\begin{figure*}
\center\includegraphics[width=0.9\textwidth]{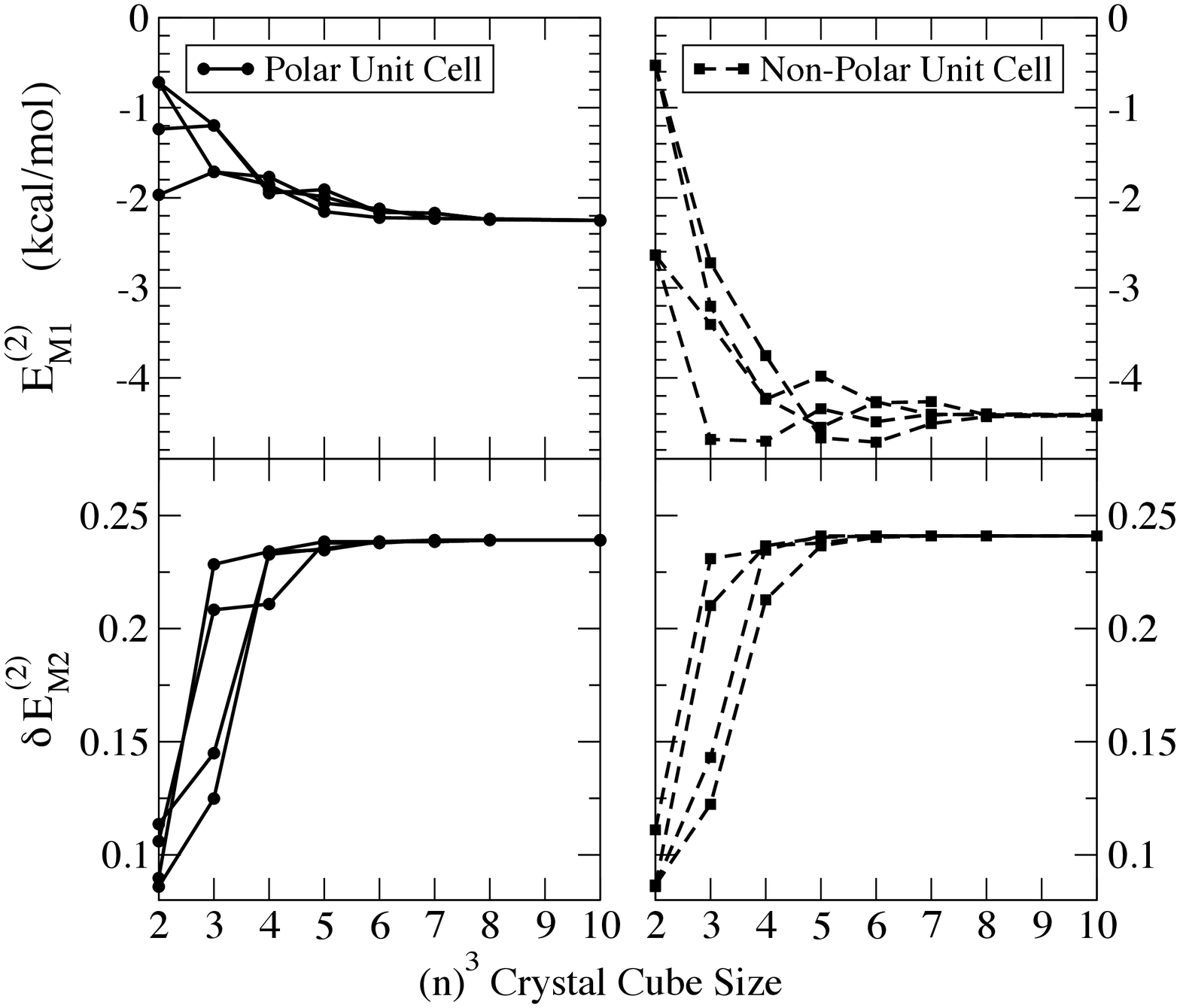}
\caption{\label{hfenergyshift}Monomer energy shifts for the four member HF
molecules of the central unit cell {for increasing crystal size}.}
\end{figure*}

In Fig. \ref{hfenergyshift}, we show the shift in energy coming from the
$E^{(2)}_{\rm M1}$ and $\delta E^{(2)}_{\rm M2}$ contribution for the
four HF molecules forming the central unit cell.  Because the crystal is explicitly
computed with no periodic boundary conditions, each internal HF molecule
experiences a different environment (including any surface polarization
effects).  To obtain true bulk limit quantities it is necessary to increase the
crystal size until interactions between the central unit cells become uniform.
We use as a metric for convergence the energy shift of the four member HF
molecules of the central unit cell.  The energy shifts perform well in this role 
as these quantities are very sensitive to the surrounding environment.  Small
deviations due to the addition of more HF molecules to the crystal (the
invariance to adding more molecules to the system being the definition of bulk
limit) become evident on the milli- and microHartree scale, well within the
numerical convergence of our calculations.

The initial $2\times2\times2$ cube is a largely artificial construct in this
bulk limit scaling analysis.  It is included as it illustrates the expected
monomer energy shift symmetry between adjacent HF chains.  As is clear for the
following small cluster sizes (3-4 per side), the energy shift varies greatly
within the central unit cell.  This is expected; surface effects still
drastically affect the central system.  Scaling the cluster size past 5
molecules per side demonstrates that the $\delta E^{(2)}_{\rm M2}$ energy shift
is essentially converged, with the $E^{(2)}_{\rm M1}$ contribution
nearly so.  The accelerated convergence of the $\delta E^{(2)}_{\rm M2}$ energy
contributions can be understood from the fact that this term is dominated by
dispersion- and induction-type interactions, which drop off as $1/R^6$ with the
intermolecular distance.  Thus contributions from anything beyond nearest
neighbors are necessarily going to be small.  This is contrasted by the 
$E^{(2)}_{\rm M1}$ contribution which is dominated by long-range dipole type
interactions that do not fully converge until at least 7 molecules per side.  We
find the $E^{(2)}_{\rm M1}$ energy shift converges for both the
$10\times10\times10$ polar and non-polar cube with an energy variance within the
unit cell of less than a micro-Hartree.  

With the perturbed HF molecular wavefunction on hand, it is straightforward to
compute the first-order response properties. In this work, we examine the dipole
moment, via using the first term of Eq. \ref{1pprop}.  Computing the dipole
moment for each HF molecule in the lattice produces an imperceptible change from
the gas phase dipole moment of 1.90 Debye.  This result is expected as the idea
of a local monomer density within a polar crystal is merely a theoretical
construct.  The average HF lattice site dipole moment per unit cell can still be
assigned by computing the non-local exchange contribution to the electronic
density and then assigning a portion of the density to a specific HF molecule
domain.  The non-local density is computed using Eq. \ref{1pdmexch}, which
uses the SCF exchange density between nearby molecule pairs.  Taking all these
effects into account, the computed average dipole moments within the central
unit cell are found to be 2.51 and 2.49 Debye for the polar and non-polar
structures respectively.  These are in excellent agreement with the published
values \cite{sode2010} of 2.51 and 2.47 Debye using MP2/aug-cc-pVDZ.  

Any quantity directly dependent on the monomer wavefunction (and energy),
including environmental effects, is of course obtainable within the MCPT
framework.  Monomer structures, conformational orderings, and vibrational
spectra are some examples that could be examined.  We can further illustrate the
utility of the MCPT method by computing the harmonic vibrational (stretch) modes
of the non-polar polymorph of the HF lattice.  This kind of property is readily
computed numerically by displacements of the nuclear coordinates and thus
requires no new methodological advancement.  With knowledge of the orthorhombic
$D_{2h}$ stretch modes \cite{hornig1955}, we can obtain the requisite single
point calculations by performing symmetric displacements about the F-H
equilibrium along the four specific stretch modes ($\rm
S(A_g),~S(B_{3u}),~S(B_{2u})\text{~and~}S(B_{1g})$) within the unit cell.  A
5-point grid for the finite-difference second derivative with a displacement
step size of $0.01$~\AA~was used with the approximation that the fluorine atoms
remain fixed.\footnote{Freezing the fluorine atoms is an approximation which
introduces a few \% error based on the reduced mass effect.}  The single point
energies used are constructed additively\footnote{The additive nature of the
monomer energy shifts is one of the strengths of the method.  It is possible
from this to compute the shifts from gas phase to liquid or solid phase of many
properties.} from Eqs. \ref{em1} and \ref{em2} by 
\begin{equation}\label{unitsum}
E^{(2)}_{\rm MCPT} = 
\sum_X\left(
E_{\rm SCF}(X) + E^{(2)}_{M1}(X) + E^{(2)}_{M2}(X)
\right)
\end{equation}
where the sum is over the four HF molecules in the central unit cell.
Having identified that the
monomer shift of the central unit cell in the $10\times10\times10$ lattice has
converged to the bulk limit, we use the same crystal size and lattice constants
for the frequency calculation for consistency.  The resulting stretch
frequencies are given in Table \ref{freqtable} as well as a reference LCCD
harmonic frequency calculation for the gas phase HF molecule.  We recover most
of the observed frequency shift relative to the gas phase, with our predicted
ordering and relative energies in agreement with experiment
\cite{kittelberger1967,anderson1980,pinnick1989}.  {As also observed
in prior periodic wavefunction results \cite{sode2012}}, the further lowering of
the $\rm S(A_g)$ and $\rm S(B_{3u})$ modes relative to the $\rm S(B_{2u})$ and
$\rm S(B_{1g})$ modes is not { achieved.  This comparison with the
values from Sode {\it et al.} is relevant as theirs are the only other computational
value in the literature.}
Further improvement to the absolute placement of the predicted spectra can be
expected through optimization of the lattice parameters, larger basis sets, and
other improved approximations which we leave to future work.


\begin{table}
\begin{tabular}{ll}
\hline
Isolated HF molecule & $\omega_e$ \\
\hline
LCCD/aug-cc-pCVDZ & 4121 \\
{CCSD/aug-cc-pCVDZ} & {4130} \\
Exp \cite{leroy1998} & 4138\\
\end{tabular}
\hfill \\
\begin{tabular}{lllll}
 & $\rm S(A_g)$ & $\rm S(B_{3u})$ & $\rm S(B_{2u})$ & $\rm S(B_{1g})$ \\
\hline
MCPT(2)/aug-cc-pCVDZ
& 3696 & 3724 & 3758 & 3770 \\
Infrared \cite{kittelberger1967} 
     & & 3067 & 3406 & \\
Raman \cite{anderson1980}
     & 3045 & & & 3386 \\
Raman \cite{pinnick1989}
     & 3027 & & & 3376 \\
{Periodic wavefunction} \cite{sode2012} & 3555 & 3458 & 3584 & 3570
\end{tabular}
\caption{\label{freqtable}Computed harmonic vibrational frequency for the gas
phase HF molecule and the four stretch modes of the non-polar polymorph HF
crystal (10$\times$10$\times$10 lattice).  Units are in cm$^{-1}$.}
\end{table}

\section{\label{conclusions}Conclusions}

In this work we have focussed on two specific goals: the formulation of a new
explicit molecular cluster based perturbation theory, and a program
implementation of said perturbation theory that is capable of scaling to the
bulk limit.  Our first goal has led us to develop the linearly scaling second-order molecular
cluster perturbation theory.  MCPT(2) contains infinite-order contributions to
the monomer electronic correlation energy and is infinite-order in the pairwise
dimer interaction (infinite-order by including all pairwise, and all possible
products of pairs).  We started with coupled-cluster perturbation theory with
the particle excitation rank Hamiltonian partitioning, as used by us
before \cite{bartlett2010,byrd2014-b} ,with the reference wavefunction taken to
be a product of monomer Hartree-Fock wavefunctions.  Limiting the cluster
expansion to only monomer and dimer interactions, a similarity transformation is
performed to obtain effective Hamiltonians that only operate on the monomer,
$\tilde{H}_X\in {\cal H}(X)$, and dimer, $\tilde{H}_{XY}\in {\cal H}(XY)$,
Hilbert spaces.  These effective Hamiltonians allow us to directly use the CCPT
framework, and help elucidate the pairing between cluster operators of the
surrounding cluster on that of a given monomer or dimer.

By working with at most explicit pairwise interactions within a cutoff, the prohibitive
{$O(N^6o^2v^4)$} computational scaling of a full CCSD explicit calculation is reduced to 
{$O(\tilde{N}No^2v^4)$} ($\tilde{N}$ is the maximum number of monomers possible
within the given cutoff radius).  For the benchmark crystal studied here (the
$10\times10\times10$ HF square crystal), the computation savings of going from
the full CCSD to the MCPT(2) framework is a factor of $10^{6}$, with an
additional factor of { $3$ ($\tilde{N}\sim 350$)} savings through the inclusion of
a cutoff.
Independent of the methodological choices that reduce computational cost, the
use of innovations in the computer science of the SIA model
Despite the
reduction due to pairwise interactions, the necessary arrays remain large.  The
automatic partitioning and inherent block sparsity capability provided by SIA
has allowed us to perform our largest HF crystal calculations on just 256 processors within a
standard $12$ hour queue length.

By using the MCPT(2) framework, we obtain not just monomer energies that include
interactions of the surrounding system to infinite-order, but also the specific
monomer wavefunction as well.  This opens up future investigations not just in
monomer energetic such as vibrational energy shifts or monomer conformer
orderings, but also direct computation of first and second-order properties of
the monomers.  

\section{Acknowledgements}

J.N.B., V.F.L. and R.J.B. acknowledge funding support from the United States Air
Force Office of Scientific Research grant FA 9550-11-1-0065 
N.J. and B.A.S. were supported by the National Science Foundation Grant
OCI-0725070 and the Office of Science of the U.S. Department of Energy under
grant DE-SC0002565.  Additional acknowledge must be given to the United States
Army Research Office DURIP grant W911-12-1-0365 which funded access to the
University of Florida Research Computing HiPerGator high performance cluster and
provided computer time at the ERDC DoD Supercomputer Resource Center.


\begin{thebibliography}{45}
\providecommand{\url}[1]{\texttt{#1}}
\providecommand{\urlprefix}{URL }
\markboth{Taylor \& Francis and I.T. Consultant}{Molecular Physics}

\bibitem{flocke2004}
N. Flocke and R.J. Bartlett,  J. Chem. Phys.  \textbf{121}, 10935 (2004).

\bibitem{tomasi2005}
J. Tomasi, B. Mennucci and R. Cammi,  Chem. Rev.  \textbf{105}, 2999 (2005).

\bibitem{gordon2007}
M. Gordon, L. Slipchenko, H. Li and J. Jensen,  Annu. Rep. Comput. Chem.
  \textbf{3}, 177 (2007).

\bibitem{fedorov2012}
D.G. Fedorov, T. Nagata and K. Kitaura,  Phys. Chem. Chem. Phys.  \textbf{14},
  7562 (2012).

\bibitem{klamt2011}
A. Klamt,  WIREs Comput. Mol. Sci.  \textbf{1}, 699 (2011).

\bibitem{vreven2006}
T. Vreven and K. Morokuma, in \emph{Annual Reports in Computational Chemistry},
  edited by D.~C. Spellmeyer, Vol.~2, Chap.~3  (Elsevier, Amsterdam, 2006), pp.
  35--51.

\bibitem{podeszwa2008}
R. Podeszwa, B. Rice and K. Szalewicz,  Phys. Rev. Lett.  \textbf{101} (11),
  115503 (2008).

\bibitem{mayhall2011}
N.J. Mayhall and K. Raghavachari,  J. Chem. Theory Comput.  \textbf{7}, 1336
  (2011).

\bibitem{knizia2012}
G. Knizia and G. Chan,  Phys. Rev. Lett.  \textbf{109}, 186404 (2012).

\bibitem{gordon2012}
M.S. Gordon, D.G. Fedorov, S.R. Pruitt and L.V. Slipchenko,  Chem. Rev.
  \textbf{112}, 632 (2012).

\bibitem{buckingham1967}
A.D. Buckingham,  Adv. Chem. Phys.  \textbf{12}, 107 (1967).

\bibitem{rybak1991}
S. Rybak, B. Jeziorski and K. Szalewicz,  J. Chem. Phys.  \textbf{95}, 6576
  (1991).

\bibitem{jeziorski1994}
B. Jeziorski, R. Moszynski and K. Szalewicz,  Chem. Rev.  \textbf{94}, 1887
  (1994).

\bibitem{chalasinski1977}
G. Chalasinski, B. Jeziorski and K. Szalewicz,  Int. J. Quant. Chem
  \textbf{11} (2), 247 (1977).

\bibitem{lotrich1997}
V.F. Lotrich and K. Szalewicz,  J. Chem. Phys.  \textbf{106}, 9688 (1997).

\bibitem{cvitas2007}
M.T. Cvita\v{s}, P. Sold{\'a}n, J.M. Hutson, P. Honvault and J.M. Launay,  J.
  Chem. Phys.  \textbf{127}, 074302 (2007).

\bibitem{bartlett2007}
R. Bartlett and M. Musia\l,  Rev. Mod. Phys.  \textbf{79}, 291 (2007).

\bibitem{cammi2009}
R. Cammi,  J. Chem. Phys  \textbf{131}, 164104 (2009).

\bibitem{bygrave2012}
P.J. Bygrave, N.L. Allan and F.R. Manby,  J. Chem. Phys.  \textbf{137}, 164102
  (2012).

\bibitem{list2014}
N.H. List, S. Coriani, J. Kongsted and O. Christiansen,  J. Chem. Phys.
  \textbf{141}, 244107 (2014).

\bibitem{bartlett2010}
R.J. Bartlett, M. Musial, V.F. Lotrich and T. Kus, in \emph{Recent Progress in
  Coupled-Cluster Methods}, edited by P. Carsky, J. Paldus and J. Pittner,
  Vol.~11, Chap.~1  (Springer, Dordrecht, 2010), pp. 1--34.

\bibitem{purvis1982}
G.D. {Purvis III} and R.J. Bartlett,  J. Chem. Phys.  \textbf{76}, 1910 (1982).

\bibitem{grabouski2007}
I. Grabowski, V. Lotrich and R.J. Bartlett,  J. Chem. Phys.  \textbf{127}
  (2007).

\bibitem{atoji1954}
M. Atoji and W. Lipscomb,  Acta Cryst  \textbf{7}, 173 (1954).

\bibitem{johnson1975}
M.W. Johnson, E. Sandor and E. Arzi,  Acta Crystallogr. Sect. B-Struct. Sci.
  \textbf{31}, 1998 (1975).

\bibitem{otto1986}
P. Otto and E.O. Steinborn,  Solid State Commun.  \textbf{58}, 281–284
  (1986).

\bibitem{panas1993}
I. Panas,  Int. J. Quantum Chem.  \textbf{46}, 109 (1993).

\bibitem{berski1998}
S. Berski and Z. Latajka,  J. Mol. Struct.  \textbf{450}, 259–263 (1998).

\bibitem{buth2004}
C. Buth and B. Paulus,  Chem. Phys. Lett.  \textbf{398}, 44–49 (2004).

\bibitem{buth2006}
C. Buth and B. Paulus,  Phys. Rev. B  \textbf{74}, 045122 (2006).

\bibitem{sode2010}
O. Sode and S. Hirata,  J. Phys. Chem. A  \textbf{114}, 8873 (2010).

\bibitem{kittelberger1967}
J.S. Kittelberger and D.F.J. Hornig,  Chem. Phys.  \textbf{46}, 3099 (1967).

\bibitem{anderson1980}
A. Anderson, B. Torrie and W. Tse,  Chem. Phys. Lett.  \textbf{70}, 300 (1980).

\bibitem{pinnick1989}
D. Pinnick, A. Katz and R. Hanson,  Phys. Rev. B: Condens. Matter Mater. Phys.
  \textbf{39}, 8677 (1989).

\bibitem{sode2009}
O. Sode, M. Ke\c{c}eli, S. Hirata and K. Yagi,  Int. J. Quantum Chem.
  \textbf{109}, 1928–1939 (2009).

\bibitem{sode2012}
O. Sode and S. Hirata,  Phys. Chem. Chem. Phys.  \textbf{14}, 7765 (2012).

\bibitem{shavitt2009}
I. Shavitt and R.J. Bartlett   (Cambridge, New York, 2009).

\bibitem{byrd2014-b}
J.N. Byrd, V.F. Lotrich and R.J. Bartlett,  J. Chem. Phys.  \textbf{140},
  234108 (2014).

\bibitem{lotrich2010}
V.F. Lotrich, J.M. Ponton, A.S. Perera, E. Deumens, R.J. Bartlett and B.A.
  Sanders,  Mol. Phys.  \textbf{108}, 3323 (2010).

\bibitem{woon1995}
D.E. Woon and T.H. {Dunning Jr. },  J. Chem. Phys.  \textbf{103}, 4572 (1995).

\bibitem{dunning1989}
T. {Dunning Jr},  J. Chem. Phys.  \textbf{90}, 1007 (1989).

\bibitem{halkier1999}
A. Halkier, W. Klopper, T. Helgaker and P. J{\o}rgensen,  J. Chem. Phys.
  \textbf{111}, 4424 (1999).

\bibitem{kendall1992}
R. Kendall, T. {Dunning Jr} and R. Harrison,  J. Chem. Phys.  \textbf{96}, 6796
  (1992).

\bibitem{hornig1955}
D.F. Hornig and W.E. Osberg,  J. Chem. Phys.  \textbf{23}, 662 (1955).

\bibitem{leroy1998}
R.J. Le~Roy,  J. Mol. Spec.  \textbf{194}, 189–196 (1999).

\end{thebibliography}


\clearpage

\clearpage

\clearpage

\clearpage

\clearpage

\end{document}